\documentstyle[preprint,amsfonts,aps,epsf]{revtex}

\begin{document}
\draft
\title{Orientational order in dipolar fluids consisting of 
nonspherical hard particles}
\author{B. Groh and S. Dietrich}
\address{
  Fachbereich Physik, Bergische Universit\"at Wuppertal, \\
 D--42097 Wuppertal, Federal Republic of Germany}

\maketitle

\begin{abstract}

We investigate fluids of dipolar hard particles by a certain variant
of density-functional theory. The proper treatment of the long range
of the dipolar interactions yields a contribution to the free energy
which favors ferromagnetic order. This corrects previous theoretical
analyses. We determine phase diagrams for dipolar ellipsoids and
spherocylinders as a function of the aspect ratio of the particles and
their dipole moment. In the nonpolar limit the results for the phase
boundary between the isotropic and nematic phase agree well with
simulation data. Adding a longitudinal dipole moment favors the nematic
phase. For oblate or slightly elongated particles we find a
ferromagnetic liquid phase, which has also been detected in computer
simulations of fluids consisting of spherical dipolar particles. The
detailed structure of the phase diagram and its evolution upon
changing the aspect ratio are discussed in detail.

\end{abstract}

\bigskip

\section{Introduction}

There are two basic molecular properties which can cause long-ranged
orientational order in fluids. First, as has already been shown by
Onsager \cite{Onsager}, particles of sufficiently {\it anisotropic shape}, e.g. long
rods or flat discs, form a nematic phase at high densities. This phase
transition can be induced by purely steric interactions \cite{LesHouches,TarazonaRev,AllenRev}. This has
been confirmed by computer simulations of hard ellipsoids \cite{FrenkelEll},
spherocylinders \cite{FrenkelNat,Frenkel88,Veerman,McGrotherHSC}, and
cut spheres \cite{FrenkelCS,VeermanCS}, which have become standard models of
liquid crystals. Some of these systems exhibit further transitions to a smectic
or a columnar phase. Second, there is growing evidence
that a ferromagnetically ordered nematic phase can be stabilized by
{\it dipolar interactions} between spherical particles, i.e. in the
absence of anisotropic steric interactions. This phase has
been observed in Monte Carlo simulations of dipolar soft
\cite{Patey1,Patey2} and hard \cite{Weis1,Weis2,Grest2}
spheres as well as in Stockmayer fluids \cite{Grest}. It has also been
analyzed by density-functional theory \cite{Patey3,Letter,Paper}. Due
to the long range of the dipolar interactions in this phase the
equilibrium configuration exhibits a spatially inhomogeneous
magnetization \cite{Domains},
similar to the domain formation in solid ferromagnets.

Molecules typically possess both a shape anisotropy and a
permanent dipole moment. Therefore it is interesting  to analyze the
relative importance of these two properties with respect to  the formation of
orientationally ordered phases and the crossover from a ferromagnetic
to a purely nematic phase. To this end in the present work we study the
models of dipolar hard ellipsoids and spherocylinders which cover the
models of
dipolar hard spheres as well as of nonpolar elongated or oblate hard
particles as limiting cases. These models have already been examined
by Onsager's virial expansion \cite{Griechen}, integral equation
theories \cite{PateyEll}, and different kinds of
density-functional theory \cite{BausColot,Vega}. However, some of these
approaches \cite{Griechen,PateyEll,Vega} suffer from an incorrect treatment of
the long-ranged dipolar forces and therefore fail to predict a
ferromagnetic phase. In simulations of dipolar hard ellipsoids
\cite{ZLW92} and spherocylinders \cite{WLZ93,LWZ93,Weis1} and the
dipolar Gay-Berne model \cite{Satoh} this phase has also not yet been
found, probably because the simulations were restricted to large
elongations of the particles which tend to destabilize the
ferromagnetic order. Terentjev et al. \cite{Terentjev} find
theoretical indications that a ferromagnetic phase might form more
readily in dipolar liquid-crystalline polymers.

In the present work we examine the models mentioned above by an alternative
density-functional theory which is a generalization of the theory
applied to Stockmayer fluids in Refs.~\cite{Letter,Paper,Domains}.

\section{Density-functional theory} \label{DFT}

As motivated in the introduction, we consider fluids consisting of hard particles which have a symmetry axis
and carry in their center a pointlike permanent dipole moment aligned with this axis. The
orientation of these uniaxial particles with respect to spatially
fixed coordinates is described by two angles
$(\theta,\phi)=\omega$. The interaction pair potential $w({\bf r}_{12},\omega,\omega')$ is the sum of the
hard core potential and the dipolar potential: $w=w_{hc}+w_{dip}$. The former is given by
\begin{equation}
  w_{hc}({\bf r}_{12},\omega,\omega')=\left\{
  \begin{array}{ll}
    \infty, & {r_{12}}\leq\sigma(\omega_{12},\omega,\omega') \\
    0,      & \text{otherwise}
  \end{array} \right.
\end{equation}
where ${\bf r}_{12}={\bf r}-{\bf r}'$ is the center-to-center distance vector
between the two particles at ${\bf r}$ and ${\bf r}'$, respectively, and
$\sigma(\omega_{12},\omega,\omega')$ is the distance of closest
approach for given orientations $\omega$, $\omega'$, and $\omega_{12}$
of the particle axes and the vector ${\bf r}_{12}$, respectively. The
dipolar potential has the form
\begin{eqnarray} \label{wdip}
  w_{dip}({\bf r}_{12},\omega,\omega')&=&-{{m^2}\over{{r_{12}}^3}}
  \left[3 (\hat{\bf m}(\omega)\cdot\hat{\bf r}_{12})(\hat{\bf
  m}(\omega')\cdot\hat{\bf r}_{12})-\hat{\bf m}(\omega)\cdot\hat{\bf 
  m}(\omega') \right] \\
  & = & {{m^2}\over{{r_{12}}^3}} \tilde w(\omega_{12},\omega,\omega'). \nonumber
\end{eqnarray}
In Eq.~(\ref{wdip}) ${\bf m}(\omega)$ is the dipole vector and $m$ is its
absolute value. The hats denote unit vectors.

\subsection{Reference system of hard particles}

We first analyze the reference system of the corresponding
nonpolar hard core fluid. Its free energy as a functional of the
number density $\hat\rho({\bf r},\omega)$ of particles at ${\bf r}$ with
orientation $\omega$ can be written as 
\begin{equation}
  F_{ref}=F_{id}+F_{ex}
\end{equation}
where the ideal gas part is given by (here and in the following we use
the notations $\int d^3r=\int_V d^3r$ and $\int d\omega=\int_{S_2}
d\omega$ where $V$ denotes the volume of the sample and $S_2$ the unit sphere)
\begin{equation}
  \beta F_{id}=\int d^3r d\omega \hat\rho({\bf r},\omega)
[\ln(4\pi\hat\rho({\bf r},\omega) \lambda^3)-1],
\end{equation}
$\lambda$ being the thermal wavelength. The excess part of the free
energy is related to the direct correlation function
$c({\bf r},{\bf r}',\omega,\omega';[\hat\rho])$:
\begin{equation} \label{cdef}
  -{1\over{k_B T}}{{\delta^2
   F_{ex}[\hat\rho]}\over{\delta\hat\rho({\bf r},\omega)\delta\hat\rho({\bf r}',\omega')}}
  =c({\bf r},{\bf r}',\omega,\omega';[\hat\rho]).
\end{equation}
Equation (\ref{cdef}) can be integrated twice along a linear path in
density space starting from a  zero density state yielding
\cite{Colot}
\begin{equation} \label{fexallg}
  \beta F_{ex}=-\int d^3r d\omega d^3r' d\omega'
  \int_0^1 d\lambda (1-\lambda) c({\bf r},{\bf r}',\omega,\omega';[\lambda
  \hat\rho]) \hat\rho({\bf r},\omega) \hat\rho({\bf r}',\omega').
\end{equation}
Due to the absence of exact results in order to proceed one now needs an approximation for the direct
correlation function. An educated guess, which renders a  computationally
simple approach but nevertheless yields reliable results for the isotropic-nematic
transition of nonpolar hard particles, is given by the decoupling
approximation introduced by Pynn \cite{Pynn} which assumes that this anisotropic
function can be obtained from a dimensionless  function $c_0(x;\eta)$
of a single variable by a suitable anisotropic
rescaling with the distance of closest approach:
\begin{equation}
  c({\bf r},{\bf r}',\omega,\omega';[\hat\rho]) \approx
   c_0({r_{12}}/\sigma(\omega_{12},\omega,\omega');\eta).
\end{equation}
Here $\eta=\rho v^{(0)}$ is the packing fraction of the particles with
an individual
volume $v^{(0)}$ and $\rho={1\over{4\pi V}} \int d^3r d\omega
\hat\rho({\bf r},\omega)$ is the mean number density. In this work we
confine ourselves to spatially homogeneous phases  so that
$\hat\rho({\bf r},\omega)=\rho\,\alpha(\omega)$ with the normalized
orientational distribution $\alpha(\omega)$; $\int d\omega
\alpha(\omega)=1$. This implies that we do not consider smectic or
solid phases and that in the case of ferromagnetic order the sample
shape is taken to be needlelike which suppresses the formation of
domains \cite{Domains}. With these assumptions and approximations Eq.~(\ref{fexallg}) reduces to
\begin{equation} \label{fex2}
  {{\beta F_{ex}}\over V}=\rho^2 f_0(\eta) \int d\omega d\omega'
   d\omega_{12} \alpha(\omega) \alpha(\omega')
  \sigma^3(\omega_{12},\omega,\omega').
\end{equation}
The function $c_0$ enters only in the integrated form
\begin{equation}
  f_0(\eta)=-\int_0^\infty dx x^2 \int_0^1 d\lambda (1-\lambda)
  c_0(x;\lambda \eta).
\end{equation}
If the Percus-Yevick direct correlation function
\cite{Hansen} was used for $c_0$ one would end up with the Percus-Yevick result for
the free energy in the special case of hard spheres (for which obviously
$\alpha(\omega)=1/(4\pi)$). Instead, in accordance with Lee
\cite{Lee1,Lee2}, we choose 
\begin{equation}
  f_0(\eta)={1\over 24}{{4-3\eta}\over{(1-\eta)^2}}
\end{equation}
which follows from the requirement that Eq.~(\ref{fex2}) yields the
Carnahan-Starling expression \cite{CS} for the free energy of hard
spheres which is known to be more accurate at high densities. Note that $f_0(\eta)$ does not depend on the shape of the
particles which enters Eq.~(\ref{fex2}) only via the distance
$\sigma(\omega_{12},\omega,\omega')$ of closest
approach.

In view of the molecular symmetry  the orientational distribution
$\alpha(\omega)$ is expected to be axially symmetric so that
\begin{equation} \label{alpser}
  2\pi\alpha(\omega)=\bar\alpha(\cos\theta)=\sum_{l=0}^\infty \alpha_l
  P_l(\cos\theta);
\end{equation}
$P_l(x)$ are the Legendre polynomials.
The excluded volume for fixed orientations $\omega$ and $\omega'$ is given
by
\begin{equation} \label{vexser}
  v_{excl}(\omega,\omega')=v_{excl}(\cos\gamma)
  ={1\over 3} \int d\omega_{12} \sigma^3(\omega_{12},\omega,\omega')
  =\sum_{l=0}^\infty v_l P_l(\cos\gamma)
\end{equation}
where $\gamma$ denotes the angle between the directions $\omega$ and
$\omega'$. Insertion of the expansions Eqs.~(\ref{alpser}) and
(\ref{vexser}) into Eq.~(\ref{fex2}) leads to 
\begin{equation} \label{Fex}
  {{\beta F_{ex}}\over V}=3 \rho^2 f_0(\eta) \sum_{l=0}^\infty
  \left({2\over{2l+1}}\right)^2 v_l \alpha_l^2.
\end{equation}
As mentioned above this expression reduces to the 
Carnahan-Starling formula in the case of hard spheres. On the other
hand in the limit $\eta\to 0$ one recovers the first two terms of the
virial expansion  used by Onsager \cite{Onsager}. This limit is
especially helpful for very elongated particles for which the
isotropic-nematic transition occurs at very low packing
fractions.

\subsection{Dipolar interaction}

The dipolar contribution to the free energy is treated in the
so-called modified mean-field approximation
\cite{Telo,Frodl1,TeloHeisenberg},
\begin{equation} \label{Fdipallg}
  F_{dip}=-{{\rho^2}\over{2\beta}} \int d^3r d\omega d^3r' d\omega'
  \alpha(\omega) \alpha(\omega')
  \Theta({r_{12}}-\sigma(\omega_{12},\omega,\omega'))
  f_{dip}({\bf r}_{12},\omega,\omega'),
\end{equation}
with the Mayer function $f_{dip}=\exp(-\beta w_{dip})-1$ which is cut
off at contact through the Heaviside function
$\Theta({r_{12}}-\sigma(\omega_{12},\omega,\omega'))$. This expression
follows  \cite{Frodl1} from using the low-density approximation $g\approx \exp(-\beta w)$ for
the pair distribution function. The Mayer
function can be expanded as 
\begin{equation}
  f_{dip}({\bf r}_{12},\omega,\omega')=\sum_{n=1}^\infty {1\over{n!}}
  \left({{-\beta m^2}\over{{r_{12}}^3}}\right)^n \tilde
  w^n(\omega_{12},\omega,\omega').
\end{equation}
Due to the slow decay as function of ${r_{12}}$  the
first term in this series requires particular attention. For this
so-called {\it l}\/ong-{\it r}\/anged term 
\begin{equation}
  F_{dip}^{(LR)}={1\over 2}\rho^2\int d\omega d\omega' \alpha(\omega)
  \alpha(\omega') \int d^3r d^3r'
  \Theta({r_{12}}-\sigma(\omega_{12},\omega,\omega'))
  w_{dip}({\bf r}_{12},\omega,\omega')
\end{equation}
the spatial integrations have to be analyzed carefully by first
considering a fluid confined to a finite volume $V$ and then performing the
thermodynamic limit for a fixed shape of this volume. As has been shown
in Ref.~\cite{Paper} the result does depend on the shape of the  sample.
For an ellipsoidal volume of aspect ratio $k$ it was found that
\begin{equation} \label{FLRsph}
  {{F_{dip}^{(LR)}}\over V}={{8\pi}\over 9}\rho^2 m^2 \alpha_1^2
(D(k)-1/3) \qquad\text{(for spherical particles)}
\end{equation}
where $D(k)$ is the demagnetization factor (see Eqs.~(3.22) and (3.24)
in Ref.~\cite{Paper}). For nonspherical particles the
spatial integrations can be separated into contributions with ${r_{12}}\leq
R_c$ and ${r_{12}}\geq R_c$ where $R_c$ is a fixed distance larger than the maximum of
$\sigma(\omega_{12},\omega,\omega')$. Since the result in Eq.~(\ref{FLRsph})
does not depend on the particle size it can be adopted for the latter
contribution (${r_{12}}\geq R_c$). The remaining integral (${r_{12}}\leq R_c$) can easily be
evaluated since in this case the kernel is effectively short-ranged as
function of ${r_{12}}$ yielding
\begin{equation}
  {{F_{dip}^{(LR)}}\over V}={{8\pi}\over 9}\rho^2 m^2 \alpha_1^2
  (D(k)-1/3)+{{\rho^2}\over{2\beta}} \int d\omega d\omega' \alpha(\omega)
  \alpha(\omega') q^{(LR)}(\cos\gamma)
\end{equation}
with
\begin{equation}
  q^{(LR)}(\cos\gamma)=\beta m^2 \int d\omega_{12} \tilde
  w(\omega_{12},\omega,\omega') \ln{{R_c}\over{\sigma(\omega_{12},\omega,\omega')}}
  =\sum_{l=0}^\infty q_l^{(LR)} P_l(\cos\gamma).
\end{equation}
(As it should be $q^{(LR)}$ does not depend on the arbitrary parameter $R_c$ due to $\int d\omega_{12} \tilde
w(\omega_{12},\omega,\omega')=0$.) As mentioned above here we consider only
the limit of an infinitely long and thin sample ($k\to\infty$) for
which $D(k)=0$. For other sample shapes the equilibrium
configuration of a ferromagnetic fluid, i.e. with $\alpha_1\neq 0$,
exhibits an inhomogeneous structure with a spatially varying axis of
preferential orientation, analogouus to the domain formation in
solid ferromagnets. It was shown in Ref.~\cite{Domains} that if this effect is
taken into account the free energy is independent of the sample shape
(see also Ref.~\cite{Griffiths}) and has the same value as for a spatially
homogeneous liquid in the limit $k\to\infty$ and thus the same phase diagram. The other terms in
Eq.~(\ref{Fdipallg}), i.e. the {\it s}\/hort-{\it r}\/anged contributions, can be written
in the form
\begin{equation} 
  {{F_{dip}^{(SR)}}\over V}={{\rho^2}\over{2\beta}} \int d\omega d\omega'
  \alpha(\omega) \alpha(\omega') q^{(SR)}(\cos\gamma)
\end{equation}
with
\begin{equation} \label{qSRdef}
  q^{(SR)}(\cos\gamma)=\int d\omega_{12} \sum_{n=2}^\infty
   {{(-1)^{n+1}}\over{3 (n-1) n!}} {{(\beta
  m^2)^n}\over{\sigma^{3n-3}(\omega_{12},\omega,\omega')}} 
  \tilde w^n(\omega_{12},\omega,\omega').
\end{equation}
With the definition
\begin{equation} \label{qldef}
  q_l={{2l+1}\over2}\int_{-1}^1 d(\cos\gamma) P_l(\cos\gamma)
 \left(q^{(LR)}(\cos\gamma)+q^{(SR)}(\cos\gamma)\right)
  -\delta_{l,1} {{4\pi}\over 3} \beta m^2
\end{equation}
($ D(k=\infty)=0$) one finally has
\begin{equation} \label{Fdip}
  {{\beta F_{dip}}\over V}={1\over 2}\rho^2\sum_{l=0}^\infty
  \left({2\over{2l+1}}\right)^2 q_l \alpha_l^2.
\end{equation}

\subsection{Total free energy and phase coexistence}
Since Eq.~(\ref{Fdip}) has the same form as Eq.~(\ref{Fex}) we can write the
total free energy as in Ref.~\cite{Paper}:
\begin{equation} \label{Fges}
  {F\over V}={\rho\over\beta} [\ln(\rho\lambda^3)-1]
  +{\rho\over\beta} \int_{-1}^1 dx\,\bar\alpha(x) \ln(2\bar\alpha(x))
  +\rho^2\sum_{l=0}^\infty u_l \alpha_l^2
\end{equation}
with (now density-dependent) coefficients
\begin{equation}
  \beta u_l=\left({2\over{2l+1}}\right)^2 (3f_0(\eta) v_l +{1\over 2} q_l).
\end{equation}
If the summation over $l$ is truncated at $l=L$ (in practice it turned
out that $L=4$ is sufficient to yield reliable results) the minimization
 with respect to the orientational distribution
leads to 
\begin{equation} \label{alqform}
   \bar\alpha(x)\sim \exp\left(-\rho\beta \sum_{i=1}^L u_i \alpha_i
   P_i(x)\right)
\end{equation}
so that (see Eq.~(5.6) in Ref.~\cite{Paper})
\begin{equation} \label{aleq}
  \alpha_l={{2l+1}\over 2} {{\int_{-1}^1 dx\,P_l(x)
  \exp\left(-\rho\beta \sum_{i=1}^L (2i+1) u_i \alpha_i P_i(x)\right)}
  \over {\int_{-1}^1 dx\,
  \exp\left(-\rho\beta \sum_{i=1}^L (2i+1) u_i \alpha_i P_i(x)\right)}}.
\end{equation}

The phase diagrams follow from requiring the equality of the chemical
potentials and the pressures at the coexisting densities $\rho_1$ and
$\rho_2$:
\begin{eqnarray}
  \left.{{\partial F}\over{\partial\rho}}\right|_{\rho_1,\bar\alpha^{(1)}(x)}&=&
  \left.{{\partial F}\over{\partial\rho}}\right|_{\rho_2,\bar\alpha^{(2)}(x)}
  \\
  F(\rho_1,\bar\alpha^{(1)}(x))-\rho_1 \left.{{\partial
  F}\over{\partial\rho}}\right|_{\rho_1,\bar\alpha^{(1)}(x)}&=&F(\rho_2,\bar\alpha^{(2)}(x))-\rho_2
  \left.{{\partial
    F}\over{\partial\rho}}\right|_{\rho_2,\bar\alpha^{(2)}(x)}. \nonumber
\end{eqnarray}
The functions $\bar\alpha^{(i)}(x)$ denote the corresponding
equilibrium orientational distributions obtained from
Eqs.~(\ref{alqform}) and (\ref{aleq}). As discussed above three kinds of phases are
considered: isotropic liquid (or gas) ($\bar\alpha(x)=1/2$), nematic liquid
($\bar\alpha(x)=\bar\alpha(-x)$, i.e. $\alpha_l=0$ for odd $l$), and ferromagnetic
liquid ($\alpha_l\neq 0$ for all $l$). (The latter phase could also be
called ferromagnetic nematic, but we do not use this phrase in order to avoid
confusion.) 

The determination of the phase boundaries for {\it second} order phase
transitions is presented in the Appendix. Note that due to symmetry
reasons there can be no truly second order isotropic-nematic transition
\cite{Chandrasekhar}.

We now discuss the calculation of the coefficients $v_l$ and $q_l$
for the two types of particles  we are interested in, i.e., hard
spherocylinders and hard ellipsoids.
A spherocylinder consists of a cylinder of length $L$ and diameter $D$
with two hemispherical caps of the same diameter $D$. The excluded
volume is given by (see, e.g., Ref.~\cite{Griechen})
\begin{equation}
  v_{excl}(\cos\gamma)=2 D L^2 \sin\gamma+2\pi D^2 L+{{4\pi}\over 3}
D^3
\end{equation}
which leads to 
\begin{equation}
   v_0={\pi\over 2}DL^2+2\pi D^2 L+{{4\pi}\over3} D^3,
   \qquad v_2=-{{5\pi}\over 16} DL^2,
   \qquad v_4=-{{9\pi}\over 128} DL^2.
\end{equation}
(Higher order terms are neglected.)
The odd coefficients vanish due to the presence of the symmetry plane of the
(nonpolar) particles. In order to determine the coefficients $q_l$
one needs the function $\sigma(\omega_{12},\omega,\omega')$. It can be
inferred from the observation that the surface of a spherocylinder is
the set of all points with distance $D/2$ from the line segment
connecting the centers of the caps. Thus when two spherocylinders are
in contact these line segments always have the distance $D$ from each
other. It is difficult to give a closed formula for
$\sigma(\omega_{12},\omega,\omega')$ because several cases have to
be distinguished. Nonetheless a numerical code for the calculation of
this function can easily be implemented.  An algorithm for the closely
related problem of the distance between two spherocylinders with given
positions and orientations has been derived explicitly by Allen et
al. \cite{AllenRev}.

For hard ellipsoids with two equal axes of length $\sigma_\perp$ and
one axis of length $\sigma_\parallel$ Berne and Pechukas \cite{Berne}
introduced the often used approximation \cite{Colot}
\begin{equation} \label{sigmaell}
  \sigma(\omega_{12},\omega,\omega')=\sigma_\perp \left[1-\chi
  {{\cos^2\theta+\cos^2\theta'-2\chi\cos\theta\cos\theta'\cos\gamma}\over
   {1-\chi^2\cos^2\gamma}}\right]^{-1/2}
\end{equation}
where $\gamma$, $\theta$, and $\theta'$ are the angles between the directions
$\omega$ and $\omega'$, $\omega$ and $\omega_{12}$, and $\omega'$ and
$\omega_{12}$, respectively, and
\begin{equation}
  \chi={{\sigma_\parallel^2-\sigma_\perp^2}\over{\sigma_\parallel^2+\sigma_\perp^2}}
  ={{\kappa^2-1}\over{\kappa^2+1}};
\end{equation}
$\kappa=\sigma_\parallel/\sigma_\perp$.
This approximation is obtained by considering the overlap of two
ellipsoidal Gauss distributions. From Eq.~(\ref{sigmaell}) one finds
\begin{equation}
  v_{excl}(\cos\gamma)={{4\pi}\over 3} \sigma_\parallel \sigma_\perp^2
  \left({{1-\chi^2 \cos^2\gamma}\over{1-\chi^2}}\right)^{1/2}.
\end{equation}
These formulas enable one to determine the coefficients $v_l$
analytically and $q_l$ numerically. We have truncated the sum in
Eq.~(\ref{qSRdef}) at $n=30$ and found no significant changes upon including further
terms  for all considered values of the parameters $\beta
m$, $\sigma_\parallel/\sigma_\perp$, and $D/L$.

\section{Discussion of the phase diagrams}

\subsection{Non-polar ellipsoids} \label{nonpolell}

Figure~\ref{figmf0} displays the phase diagram for {\it nonpolar}
hard ellipsoids with aspect ratio
$\kappa=\sigma_\parallel/\sigma_\perp$. The solid lines denote the
coexistence densities of the isotropic and the nematic fluid as
determined from the theory presented in Sec.~\ref{DFT}. They are in
fair agreement with the results of Monte Carlo simulations (squares)
\cite{FrenkelEll}. The two-phase region is always very narrow, as in
real nematic liquid crystals. The triangles indicate the liquid-solid
transition found in the simulations, which cannot be described by the
present theory. (The description of the solid phase requires a
weighted-density-functional theory \cite{Solid}.) For aspect ratios near $\kappa=1$ the isotropic-nematic
transition is preempted by freezing. Within the approximation in
Eq.~(\ref{sigmaell}) the physical properties are invariant under the
transformation $\kappa\to 1/\kappa$. This behavior is satisfied
very well by the simulation results, too.

\subsection{Polar ellipsoids} \label{polell}

In this subsection we discuss the case that the hard ellipsoids are
endowed with a point dipole of strength $m$ oriented along the
symmetry axis of length $\sigma_\parallel$.
In the
following we use the dimensionless reduced temperature $T^\ast=k_B T
\sigma_\perp^3/m^2$ and the volume fraction $\eta=\rho v^{(0)}$ where
$v^{(0)}={{\pi}\over 6}\sigma_\perp^2 \sigma_\parallel$ is the
molecular volume. The nonpolar case (see Subsec.~\ref{nonpolell} and
Fig.~\ref{figmf0}) corresponds to the limit $T^\ast\to\infty$. For
$\kappa=3$ (Fig.~\ref{figk3}) the first-order isotropic-nematic
transition is shifted to lower densities upon lowering the
temperature. Below $T^\ast\approx0.2$ the density gap increases
dramatically and evolves into a broad coexistence region between an
isotropic gas and a nematic liquid. Thus in this fluid gas-liquid
coexistence is not terminated by a critical point. For $\eta\geq
0.265$ lowering the temperature leads to two phase transitions
(isotropic fluid$\to$nematic fluid and nematic fluid$\to$gas-liquid)
whereas for $\eta\leq 0.265$ there is only the gas-liquid transition. 
As shown in
Fig.~\ref{figk2} for $\kappa=2$ a ferromagnetic liquid appears in the
medium temperature range. This phase turns into a purely nematic phase
along the dotted lines of critical points. Whether the disappearence
of the ferromagnetic order at low temperatures is an artefact of the
approximations or not needs to be checked by alternative
techniques. The behavior of the order parameters $\alpha_l$ along
different thermodynamic paths is displayed in Figs.~\ref{figalpha} and
\ref{figalprho}. Figure~\ref{figalpha} illustrates their density
dependence along two isotherms. For $T^\ast=1.3$ the ferromagnetic order
parameter $\alpha_1$ vanishes at the critical density with a square
root singularity in accordance with the presently used mean-field
theory while the nematic order parameter $\alpha_2$ exhibits a small
break of slope not visible on the scale of the figure. At a lower density $\alpha_2$
vanishes discontinuously at 
the first-order nematic-isotropic transition. For the lower
temperature $T^\ast=0.6$ the ferromagnetic phase transforms directly into
the isotropic phase. Figure~\ref{figalprho} shows the order parameters
as function of $T^\ast$ at the fixed density $\eta=0.75$ demonstrating
the loss of the ferromagnetic order at low temperatures. In the
intermediate ferromagnetic region the orientational distribution
$\bar\alpha(\cos\theta)$ typically exhibits two maxima at $\theta=0$ and
$\theta=\pi$ with the higher one determining the sign of the
spontaneous magnetization. Their heights become equal at the
nematic-ferromagnetic transition. In contrast to Baus and Colot
\cite{BausColot}, who use a different density-functional theory, we do
not observe a phase transition between phases with one and two maxima
in the orientational distribution.

Upon further lowering the aspect ratio (Fig.~\ref{figk15}) another
qualitative change of the phase diagram occurs: for intermediate densities the
isotropic-ferromagnetic transition becomes continuous. Two
tricritical points arise where the character of this transition
changes from first to second order. We remark that for high
temperatures the ordered phases occur only at such high densities that
they will certainly be preempted by a solid phase which is not
captured by the present form of the density functional.

Finally in Fig.~\ref{figk1} we  present the phase diagram for dipolar
hard {\it spheres} obtained from the present density-functional
theory. There is no gas-liquid transition between isotropic fluids,
but a coexistence of an isotropic gas with a ferromagnetic liquid
which at a tricritical point $(T^\ast_t,\eta_t)=(0.600,0.198)$ changes into a continuous transition. As
for the Stockmayer fluid \cite{Paper} the stability of the
ferromagnetic phase is considerably overestimated as compared with
simulations \cite{Weis1,Weis2} in which the isotropic-ferromagnetic transition has been detected
at $T^\ast=0.16$ and $\eta\approx0.4$. 

For oblate particles ($\kappa<1$) we obtain a similar series of phase
diagrams (Figs.~\ref{figk23} and \ref{figk13}) as for $\kappa>1$ but
without the loss of ferromagnetic order and the reentrance of the
nematic phase at low temperatures. The comparison between Figs.~\ref{figk23}
and \ref{figk15} and between Figs.~\ref{figk13}
and \ref{figk3} shows that the formation of the ferromagnetic phase is
favored by oblate particles as compared with elongated ones. This is confirmed in Fig.~\ref{figmf1}
which displays the phase diagram in the $(\kappa,\eta)$-plane for a fixed value of
the squared dipole moment per $k_B T$ normalized to the particle volume: ${{\beta
m^2}/v^{(0)}}=6/\pi$. The isotropic-ferromagnetic critical
density increases with increasing elongations.  An intuitive
explanation for this observation is provided by an examination of the
lowest energy configuration of two dipolar particles. The interaction
energy at contact of a nose-to-tail arrangement is
$-2m^2/(\kappa\sigma_\perp)^3$ while that of an antiparallel side-by-side
arrangement is $-m^2/\sigma_\perp^3$. Obviously the former
configuration becomes more favorable as compared with the latter one if $\kappa$ is
decreased implying a stronger tendency for long-ranged ferromagnetic
order. For large elongations the orientationally ordered phase becomes
nematic while this does not happen for very oblate particles, for which
instead a gas-ferromagnetic coexistence occurs at the temperature
considered. Thus the $\kappa\leftrightarrow 1/\kappa$ symmetry of
Fig.~\ref{figmf0} is lost in Fig.~\ref{figmf1} as a result of the additional
dipolar forces.

\subsection{Dipolar spherocylinders}

For dipolar {\it spherocylinders} one finds the same series of
phase diagrams as function of the aspect ratio $L/D$ as for elongated
ellipsoids. An example is given in Fig.~\ref{figddl1} (compare
Fig.~\ref{figk2}). As for the nonpolar ellipsoids the location of the isotropic-nematic
transition for $m=0$ agrees very well with corresponding simulation results
\cite{FrenkelNat,Veerman} which are available only for $L/D=5$.

\subsection{Comparison with previous results}

In the following we compare our results with other theoretical
investigations and numerical
simulations of dipolar hard core particles. Baus and Colot
\cite{BausColot} have studied this problem with a different
density-functional ansatz which utilizes the
analytically known correlation function of dipolar hard spheres in the
mean-spherical approximation. At high temperatures they also find the
sequence of isotropic, nematic, and ferromagnetic phases in accordance
with our results, followed by a transition to a second ferromagnetic
phase (see the end of the first paragraph in
Subsec.~\ref{polell}). However, they claim that an increase of the aspect
ratio shifts the ferromagnetic phase to lower densities and higher
temperatures in contrast to our findings and to the qualitative
argument presented at the end of Subsec.~\ref{polell}.

The familiar second order virial expansion of Onsager \cite{Onsager}
has been applied to dipolar spherocylinders by Vanakaras and Photinos
\cite{Griechen}. It is well known that this approximation deteriorates for
decreasing aspect ratios \cite{Lee1} which is the reason that for $L/D=5$ and $m=0$ these
authors find the isotropic-nematic transition at densities above the
density of closest packing. In addition they do
not treat the long-ranged dipolar interactions correctly which leads
them to the wrong conclusion that a ferromagnetic phase cannot be
stable for any dipole moment or aspect ratio. They agree with our
results in that an increase of $m$ or a decrease of $T$ lowers the
isotropic-nematic coexistence densities for point dipoles located in
the center of the hard particles.
However, as they point out, this trend is reversed if the point
dipoles are located off center near the ends of the particles.

The correlation functions obtained from the  hypernetted-chain
integral equation theory have been used by Perera and Patey
\cite{PateyEll} for the investigation of dipolar hard ellipsoids. For $\kappa=3$
they find transition densities which are about 20\% lower than the
ones presented here and the same
behavior of these densities upon increasing the dipolar strength. However, for $\kappa=1/3$
these authors were able to obtain a solution of the integral equation only for rather small dipole
moments. They, too, did not take into account the correct long-ranged contribution to
the free energy and did not detect a ferromagnetic phase for the
aspect ratios under consideration. However, in a subsequent publication \cite{Patey3} they
state that this conclusion remains unaltered if this fault is corrected.

Vega and Lago \cite{Vega} have used a density-functional theory
similar to the present one, but they incorporate a more sophisticated
equation of state for the nonpolar isotropic fluid and treat the
dipolar interactions within perturbation theory. The same remark as
above concerning the treatment of the long-ranged interaction applies
also to their work. Their results for the isotropic-nematic
coexistence densities of dipolar hard spherocylinders with aspect
ratio $L/D=5$ are in fair agreement with our findings for $\beta
m^2/D^3$ between 0 and 4. However, they were unable to find a solution
for the orientational distribution function for $\beta m^2/D^3=6$
which is close to the parameter range where within our approximation
the isotropic-nematic transition broadens into a gas-nematic
coexistence.

Weis, Levesque, and Zarragoicoechea \cite{WLZ93,LWZ93,Weis1} have
performed Monte Carlo simulations of dipolar spherocylinders with
$L/D=5$ for centered and off-centered dipole moments forming different
angles with the cylinder axis. These authors were mainly interested in the
structure of the smectic phase and did not determine the
isotropic-nematic transition point. For $(\beta
m^2/D^3)^{1/2}=2.449$ and a longitudinal centered dipole
moment they report \cite{WLZ93} the occurrence of an {\it i}\/sotropic state at $\eta=0.356$ and
a {\it n}\/ematic state at $\eta=0.441$ which is in accordance with the
coexistence densities $\eta_I=0.356$ and $\eta_N=0.375$ obtained from
the present theory. These authors have also studied ellipsoids with $\kappa=3$
\cite{ZLW92} and found no isotropic-nematic transition for values of $(\beta
m^2/\sigma_\perp^3)^{1/2}$ in the range between 0 and 3,  in contrast to our results shown in
Fig.~\ref{figk3}. For
$m=0$ this finding is also at variance with Frenkel and Mulder's results
\cite{FrenkelEll}, who did observe an isotropic-nematic transition with
$\eta_I=0.507$ and $\eta_N=0.517$. We conclude that further simulations of the
isotropic-nematic or possibly the isotropic-ferromagnetic transition for different
aspect ratios would certainly be helpful. 

Recently McGrother and
Jackson \cite{McGrother} have published an extensive Monte Carlo study
of the liquid-vapor coexistence in a dipolar hard spherocylinder
fluid. In agreement with previous work \cite{Smit,Grest} they find
that there is no liquid-vapor coexistence in the spherical limit
$L/D\to 0$ as well as for very elongated particles $L/D\gg 1$ due the formation of chains of nose-to-tail
(small $L/D$) or side-by-side (large $L/D$) aggregated particles (see
also Ref.~\cite{RoijLett}). Only in
an intermediate range around $L/D=0.25$ phase separation into
isotropic gas and liquid does
occur. This chain formation is not captured correctly by the
present theory which predicts gas-ferromagnetic or gas-nematic 
coexistence for all values of $L/D$.

\section{Summary}

By applying density-functional theory for the description of
orientational order in dipolar fluids consisting of hard nonspherical
particles we have obtained the following main results:

\begin{enumerate}
\item The location of the isotropic-nematic transition of nonpolar
hard particles as a function of the aspect ratio has been obtained in good agreement
with other theories and simulation results (Fig.~\ref{figmf0}).

\item The addition of a longitudinal point dipole at the centers of the
particles induces a decrease of the coexisting densities of the
 isotropic-nematic transition and it leads to gas-nematic coexistence
for large values of the
 dipole moment or at low temperatures (Fig.~\ref{figk3}).

\item There is also a ferromagnetic liquid phase provided the
particles are not too elongated and the dipole moment is sufficiently
strong. This phase is reached from the
nematic or isotropic states by continuous or weakly first-order
transitions, depending on the temperature and the particle aspect ratio.

\item In accordance with the mean-field character of
density-functional theory, at the continuous phase transitions (Curie
points) the magnetization vanishes according to a square root power
law and the nematic order parameter exhibits a small break of slope
(Figs.~\ref{figalpha} and \ref{figalprho}).

\item Within the present theory the phase diagram of dipolar hard
spheres (Fig.~\ref{figk1}) comprises an isotropic and a
ferromagnetic fluid with first (second) order phase transitions at
temperatures below (above) a tricritical point.

\item Oblate particles exhibit a similar phase behavior as elongated
ones, but a stronger tendency for the formation of the ferromagnetic
phase. In contrast to the nonpolar case the phase diagrams are not
approximately symmetric with respect to the transformation
$\kappa\leftrightarrow 1/\kappa$, where $\kappa$ is the aspect ratio
of the uniaxial ellipsoidal particles (Fig.~\ref{figmf1}).

\item For dipolar hard spherocylinders we find an analogous series of
phase diagrams as for elongated ellipsoids.

\end{enumerate}


\appendix

\section*{Critical densities}

We rewrite the contributions to those free energy in Eq.~(\ref{Fges}) which
depend on the orientational distribution as (here and in the following
all integrals over $x$ and $x'$ are to be taken over the interval $[-1,1]$)
\begin{equation} \label{Fapp}
  {{\Delta F}\over V}={\rho\over\beta} \int dx \bar\alpha(x) \ln(2\bar\alpha(x))
   +{1\over 2}\rho \int dx dx'\,\bar\alpha(x)\bar\alpha(x') K(x,x')
\end{equation}
with
\begin{equation}
  K(x,x')=\sum_{l=1}^L {{(2l+1)^2}\over 2} u_l P_l(x) P_l(x').
\end{equation}
Following van Roij et al. \cite{Roij} we use a kind of bifurcation
analysis in order to determine the critical density of the
nematic-ferromagnetic or isotropic-ferromagnetic phase transition for a given temperature.
The minimization of Eq.~(\ref{Fapp}) yields
\begin{equation} \label{minbdg}
  \ln(2\bar\alpha(x))+\beta\rho \int dx' \bar\alpha(x') K(x,x')=\nu
\end{equation}
where the Lagrange multiplier $\nu$ is determined by the
normalization $\int dx \bar\alpha(x)=1$. One considers a small ferromagnetic
perturbation $\bar\alpha_1(x)$ of a solution $\bar\alpha_0(x)$ with nematic
symmetry, i.e. $\bar\alpha_0(x)=\bar\alpha_0(-x)$. If the expansions $\bar\alpha(x)=\bar\alpha_0(x)+\epsilon
\bar\alpha_1(x)+\ldots$ and $\nu=\nu_0+\epsilon \nu_1+\ldots$ are
inserted into Eq.~(\ref{minbdg}) the term linear in $\epsilon$ gives
\begin{equation} \label{a4}
  {{\bar\alpha_1(x)}\over{\bar\alpha_0(x)}}+\beta\rho \int dx'\,\bar\alpha_1(x')
   K(x,x')=\nu_1.
\end{equation}
Integrating Eq.~(\ref{a4}) over $x$ and using the relation $\int dx
K(x,x')=0$ yields
\begin{equation} \label{minepsbdg}
  {{\bar\alpha_1(x)}\over{\bar\alpha_0(x)}}-{1\over 2} \int dx'
  {{\bar\alpha_1(x')}\over{\bar\alpha_0(x')}}
  +\beta\rho \int dx'\,\bar\alpha_1(x')   K(x,x')=0.
\end{equation}
Since both $\bar\alpha_0(x)$ and $\bar\alpha(x)$ must be of the form
$\exp(\sum_{l=0}^L \gamma_l P_l(x))$ (see Eq.~(\ref{alqform})),  with
 the nematic solution $\bar\alpha_0(x)$ containing only even indices $l$, we
make the following ansatz for the small perturbation:
\begin{equation}
  \bar\alpha_1(x)=\bar\alpha_0(x) \sum_{n=1}^{L/2} \gamma_{2n-1} P_{2n-1}(x)
\end{equation}
where terms of the order $\gamma_{2n-1} \gamma_{2m-1}$ and higher due
to the exponential form (see above) have been neglected. Due to
Eq.~(\ref{minepsbdg}) the coefficients $\gamma_l$ with odd $l$ satisfy the
equation
\begin{equation} \label{EW}
  \gamma_l+\sum_{l'=1}^{L/2} A_{ll'} \gamma_{l'}=0
\end{equation}
with
\begin{equation} \label{Adef}
  A_{ll'}={{(2l+1)^2}\over 2}\beta\rho u_l \int dx \bar\alpha_0(x) P_l(x)
  P_{l'}(x).
\end{equation}
Both the nematic solution $\bar\alpha_0(x)$ and the $L/2\times L/2$ matrix $A$ depend
on the density. The critical density is reached if one of the eigenvalues
of $A$ equals $-1$ giving rise to a nontrivial solution of
Eq.~(\ref{EW}). For $L=2$ this condition reduces to 
\begin{equation} \label{L2bdg}
  {9\over 2} \beta\rho u_1 \int dx\,\bar\alpha_0(x) x^2=-1.
\end{equation}
For $L=4$ one obtains after some algebra
\begin{equation}
  1+A_{11}+A_{33}+A_{11} A_{33} - A_{13} A_{31}=0.
\end{equation}
From these equations together with the numerical solution of Eq.~(\ref{aleq})
the nematic-ferromagnetic critical density can be determined.

For the isotropic-ferromagnetic transition the unperturbed solution is
$\bar\alpha_0(x)=1/2$ so that Eq.~(\ref{L2bdg}) reduces to the known result (see
Eq.~(7.10) in Ref.~\cite{Paper})
\begin{equation} \label{isobdgL2}
  {3\over 2} \beta \rho u_1=-1.
\end{equation}
In this case the matrix $A$ is diagonal for general $L$: $A_{ll'}=(2l+1)/2
\beta\rho u_l \delta_{l,l'}$. Thus the above bifurcation condition yields
\begin{equation} \label{isobdgallg}
  {{2l+1}\over 2} \beta\rho u_l=-1.
\end{equation}
The actual values of the coefficients $u_l$ are such that the lowest
density for which Eq.~(\ref{isobdgallg}) is satisfied always corresponds to
$l=1$, so that Eq.~(\ref{isobdgL2}) is valid for all $L$.



\begin{figure}
\caption{Phase diagram of nonpolar hard ellipsoids with two equal
axes of length $\sigma_\perp$ and one axis of length
$\sigma_\parallel$ and an aspect ratio
$\kappa=\sigma_\parallel/\sigma_\perp$. $\eta=\rho v^{(0)}$ is the
volume fraction where $\rho$ is the number density of the particles and
$v^{(0)}$ is their individual volume. The lines denote the coexisting
volume fractions at the first-order isotropic-nematic transition as
obtained from density-functional theory. The squares and triangels are
simulation results for isotropic-nematic and liquid-solid coexistence,
respectively. For the former there is good agreement with
density-functional theory. The dotted lines are guides to the eye. The
isotropic-nematic transition is accompanied by only a small density
discontinuity which decreases for $\kappa\to 1$. The present
density-functional theory is not suited to describe the freezing
transition. The simulation data support the symmetry
$\kappa\leftrightarrow 1/\kappa$ discussed in the text. }
\label{figmf0}
\end{figure}

\begin{figure}
\caption{Phase diagram of dipolar hard ellipsoids with aspect ratio
$\kappa=3$ in the temperature-density plane. The weakly first-order
isotropic-nematic transition at high temperatures broadens into 
gas-nematic coexistence at low temperatures. Here and in the following
figures two-phase coexistence regions are shaded.}
\label{figk3}
\end{figure}

\begin{figure}
\caption{The same as in Fig.~\protect\ref{figk3} but for $\kappa=2$. Besides the
isotropic (I) and nematic (N) phases a ferromagnetically ordered
liquid (F) occurs in the intermediate temperature range. The dotted
and solid lines denote second- and first-order transitions,
respectively. The density gap of the I-N transition at $\eta\simeq 0.68$
cannot be resolved on the present scale. The continuous F-N transition
at low temperatures intersects the first-order transitions at a critical
end point. The inset shows that in a narrow temperature range the high
temperature continuous N-F transition is turned into a weakly
first-order transition generating a tricritical point and an I-N-F
triple point. The two-phase coexistence regions are shaded. For
reasons of clarity this shading is omitted for I-F coexistence.}
\label{figk2}
\end{figure}

\begin{figure}
\caption{The first two orientational order parameters $\alpha_l$ (see
Eq.~(\ref{alpser})) along two isotherms in Fig.~\protect\ref{figk2} for
$\kappa=2$. The gaps between the
black dots indicate two-phase regions. The dotted vertical lines
indicate discontinuities.}
\label{figalpha}
\end{figure}

\begin{figure}
\caption{Temperature dependence of the first four orientational order
parameters $\alpha_l$ along a thermodynamic path of
fixed density in Fig.~\protect\ref{figk2} for $\kappa=2$. The order parameters
with odd indices $l$ vanish
in the low- and high-temperature nematic phases but are nonzero between
the lower critical point $T^\ast_{c1}=0.226$ and the upper critical point
$T^\ast_{c2}=1.34$ corresponding to ferromagnetic order. $\alpha_2$ and
$\alpha_4$ are nonzero for all temperatures and decrease for increasing
temperature. At $T^\ast_{c1}$ and $T^\ast_{c2}$ they exhibit a break of slope.}
\label{figalprho}
\end{figure}

\begin{figure}
\caption{Phase diagram of dipolar hard ellipsoids for $\kappa=1.5$. The line styles have the same
meaning as in Fig.~\protect\ref{figk2}. In the high temperature region the gap
between the two coexisting densities cannot be resolved on the scale of
the figure so that only a single solid line is visible. Between two
tricritical points the isotropic-ferromagnetic transition is
continuous. The two-phase coexistence region at low temperatures is shaded.}
\label{figk15}
\end{figure}

\begin{figure}
\caption{Phase diagram of dipolar hard spheres. Within
the present approximation only one isotropic
fluid and a ferromagnetic liquid are stable at any temperature. Below
(above) the tricritical point the phase transition is discontinuous (continuous).}
\label{figk1}
\end{figure}

\begin{figure}
\caption{Phase diagram of oblate dipolar ellipsoids with
$\kappa=2/3$. The meaning of the solid and dotted lines is the same as in
Fig.~\protect\ref{figk2}. In contrast to Fig.~\protect\ref{figk15} there is no reentrant nematic phase at low temperatures.}
\label{figk23}
\end{figure}

\begin{figure}
\caption{Phase diagram for oblate dipolar ellipsoids with
$\kappa=1/3$. See also the caption to Fig.~\protect\ref{figk2}.}
\label{figk13}
\end{figure}

\begin{figure}
\caption{Phase diagram of hard ellipsoids for a fixed reduced dipole
moment in units of $k_B T$ and the particle volume $v^{(0)}$. The stability of the ferromagnetic phase decreases with
increasing aspect ratio $\kappa$ until it finally turns into a nematic
phase at high densities and into an isotropic fluid at low densities. Coexistence of an isotropic gas and the
ferromagnetic liquid occurs for oblate particles. There is no
$\kappa\leftrightarrow 1/\kappa$ symmetry as in Fig.~\protect\ref{figmf0} due to
the presence of the dipolar interactions.}
\label{figmf1}
\end{figure}

\begin{figure}
\caption{The phase diagram for polar spherocylinders with an aspect
ratio $D/L=1$ exhibits a similar behavior as that for polar ellipsoids
with $\kappa=2$ (see Fig.~\protect\ref{figk2}).}
\label{figddl1}
\end{figure}

\newcommand{\pssize}[1]{\setlength{\epsfxsize}{#1}}

\newpage
\begin{minipage}[h]{16cm}
\pssize{16cm}
\epsfbox{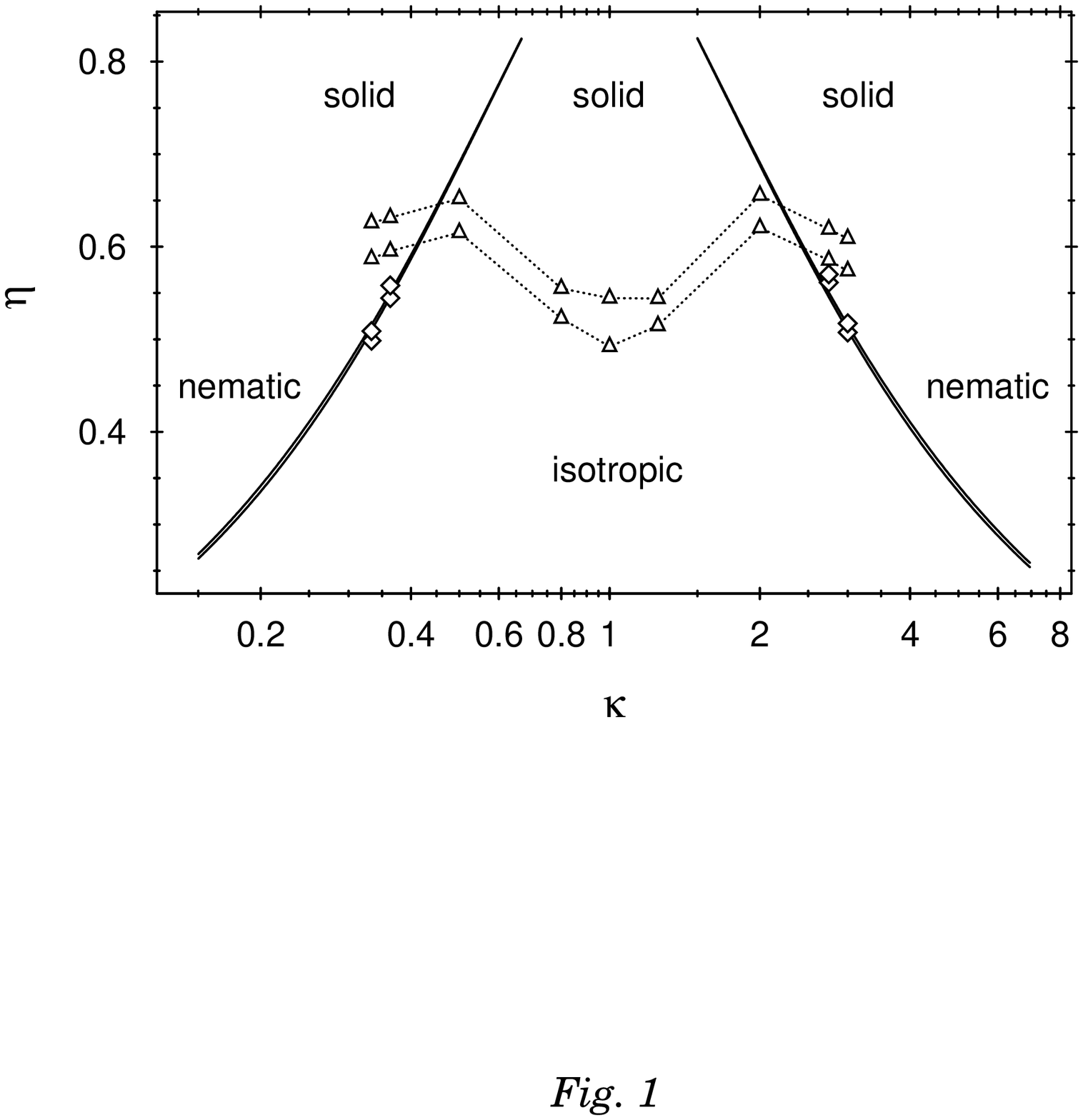}
\end{minipage}

\newpage
\begin{minipage}[h]{16cm}
\pssize{16cm}
\epsfbox{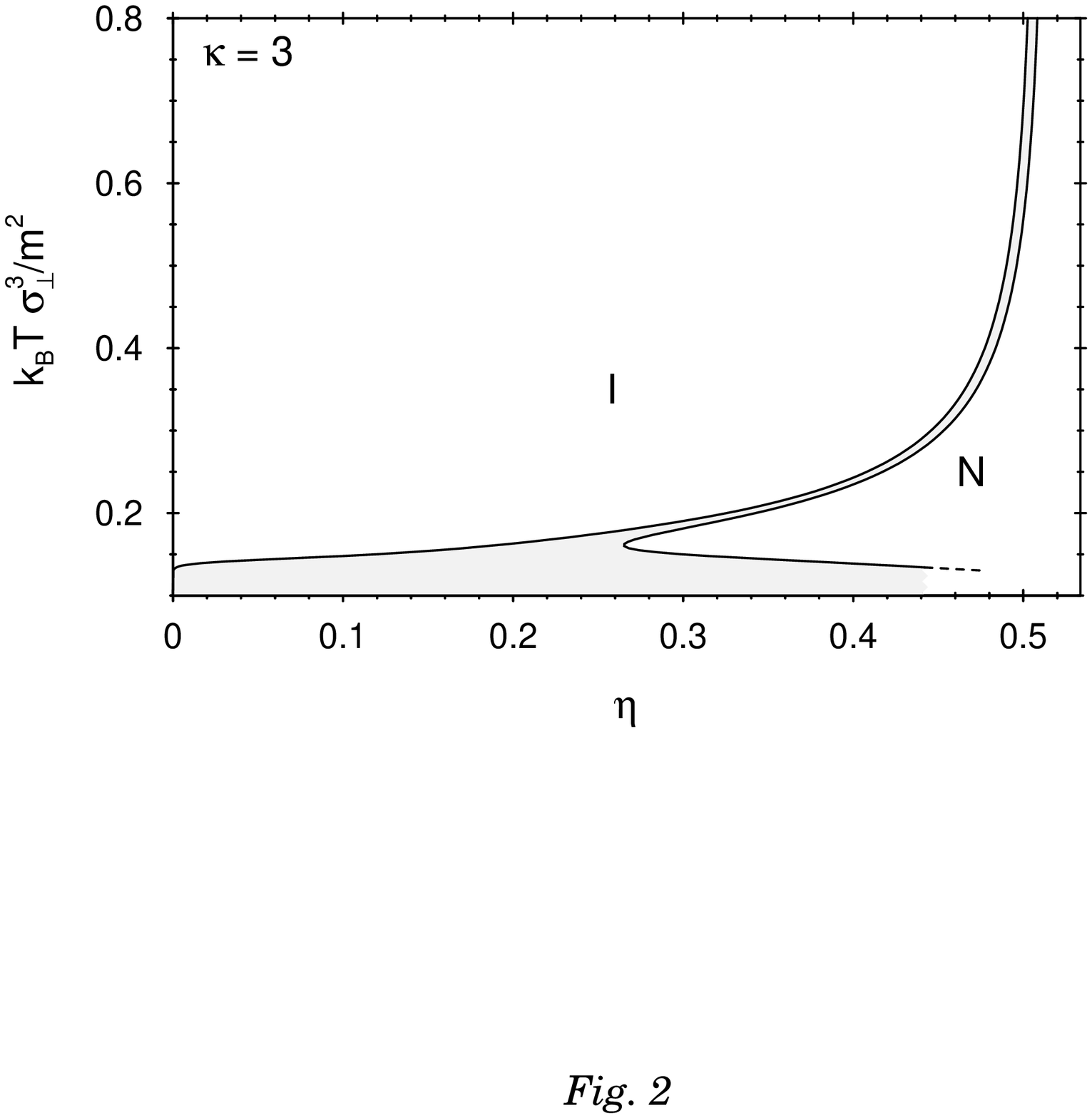}
\end{minipage}

\newpage
\begin{minipage}[h]{16cm}
\pssize{16cm}
\epsfbox{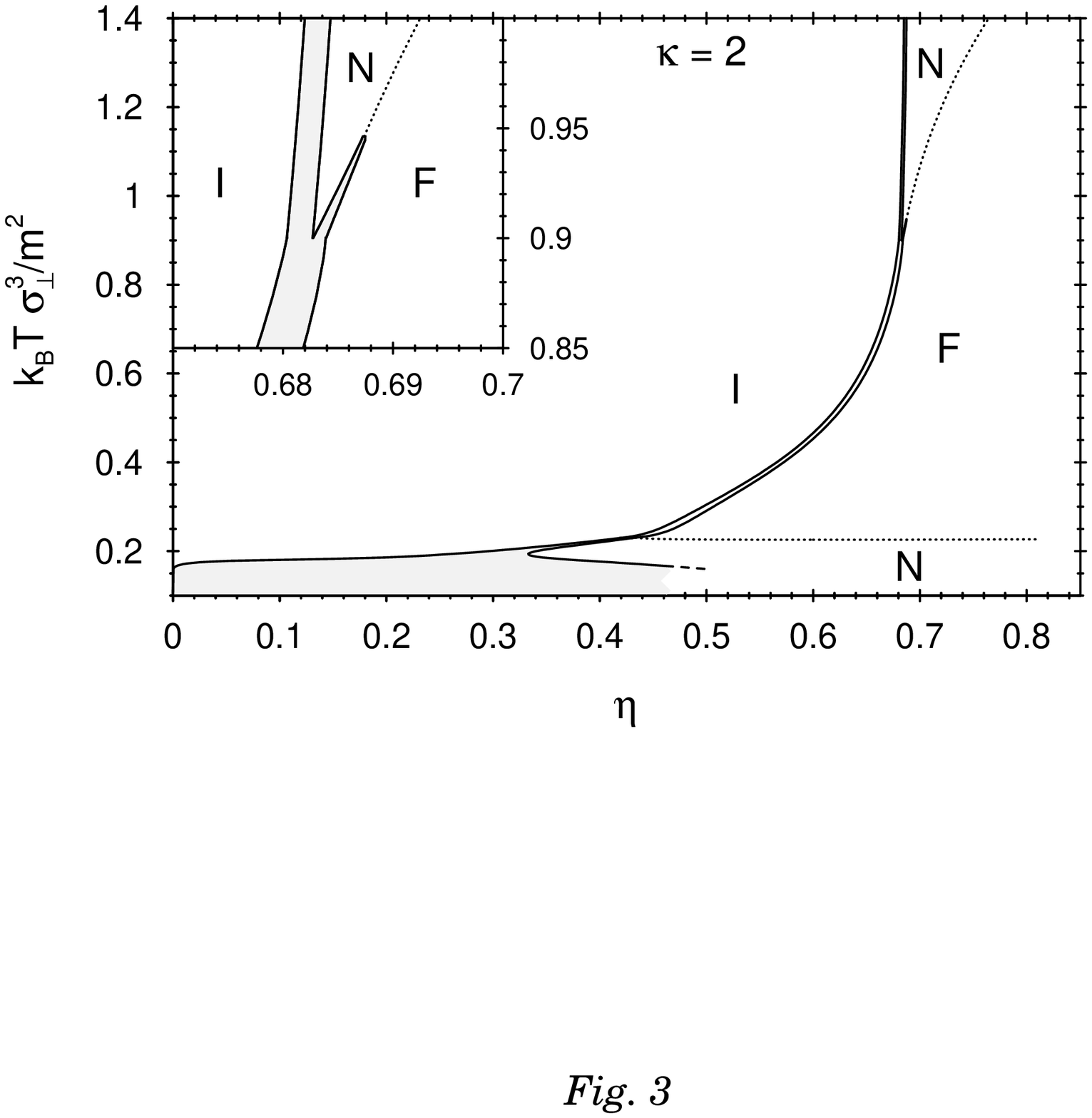}
\end{minipage}

\newpage
\begin{minipage}[h]{16cm}
\pssize{16cm}
\epsfbox{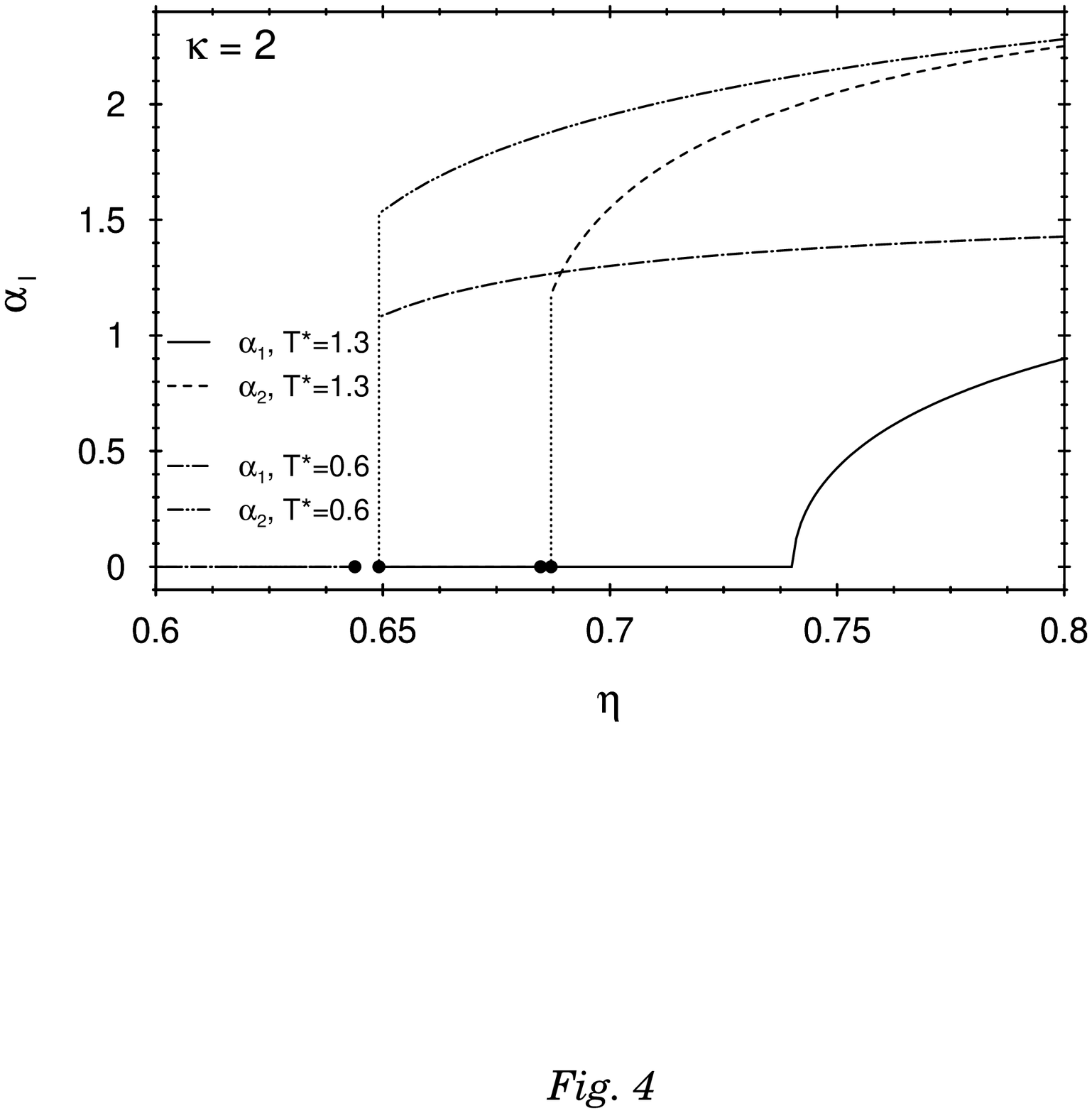}
\end{minipage}

\newpage
\begin{minipage}[h]{16cm}
\pssize{16cm}
\epsfbox{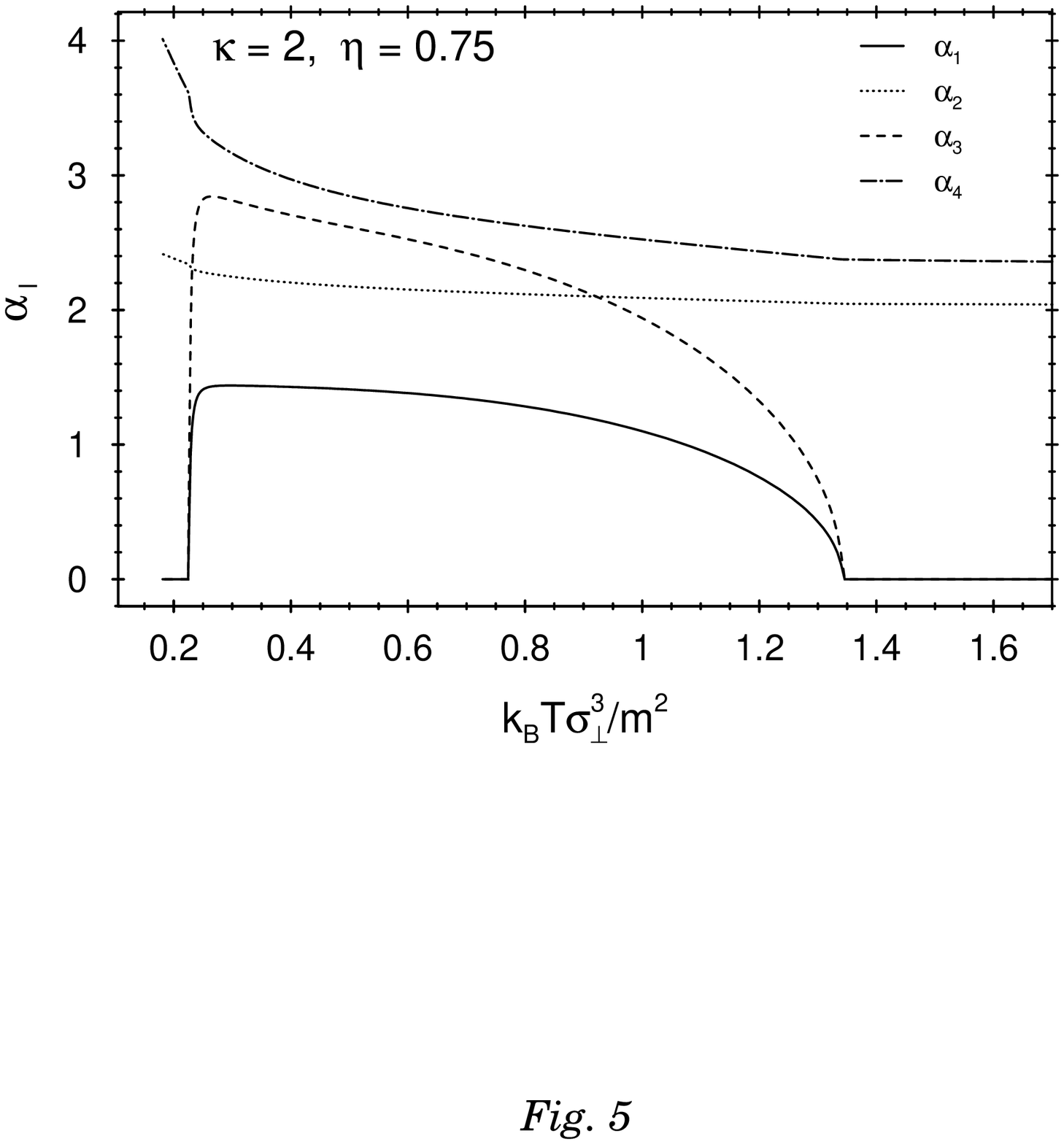}
\end{minipage}

\newpage
\begin{minipage}[h]{16cm}
\pssize{16cm}
\epsfbox{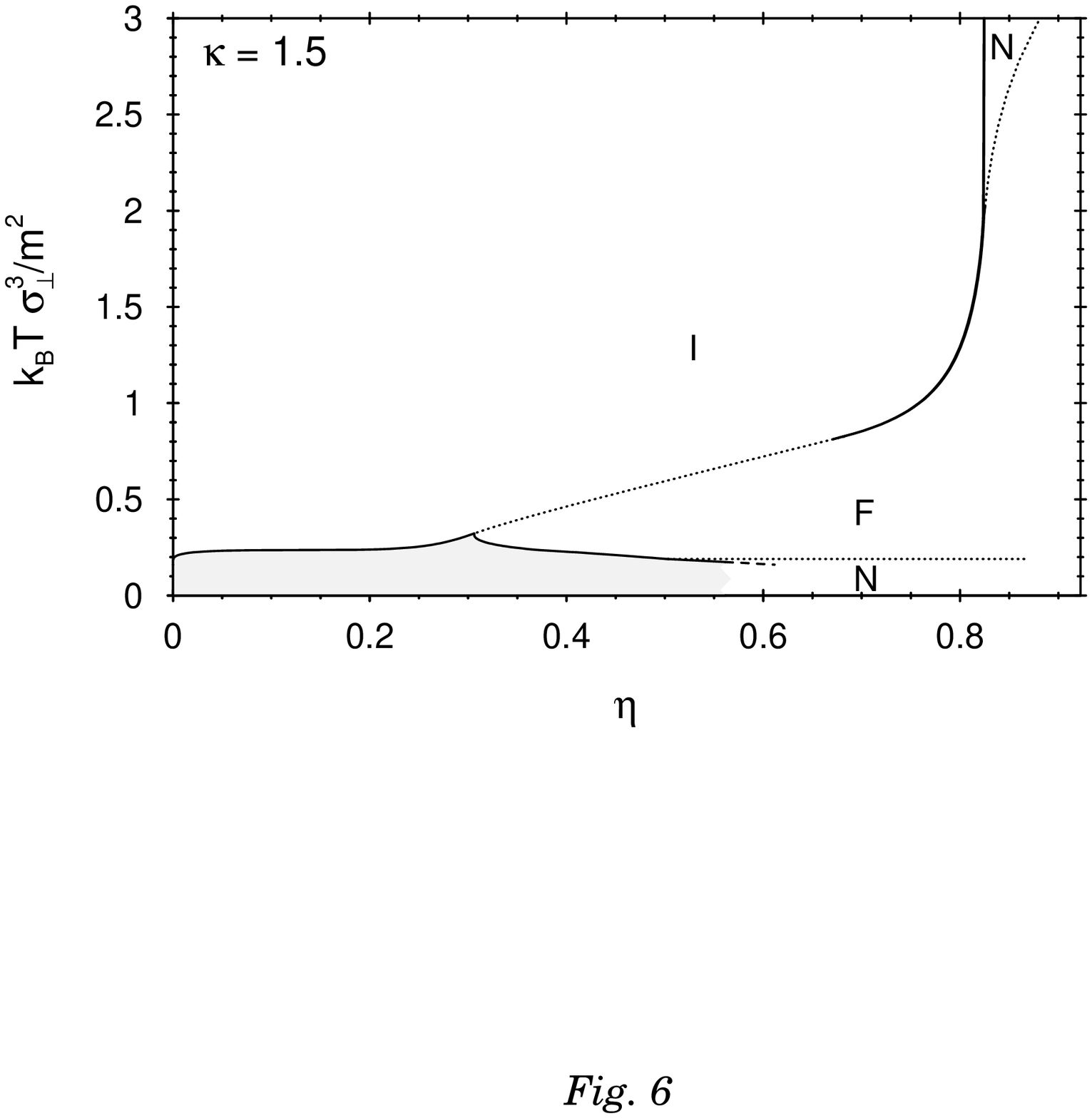}
\end{minipage}

\newpage
\begin{minipage}[h]{16cm}
\pssize{16cm}
\epsfbox{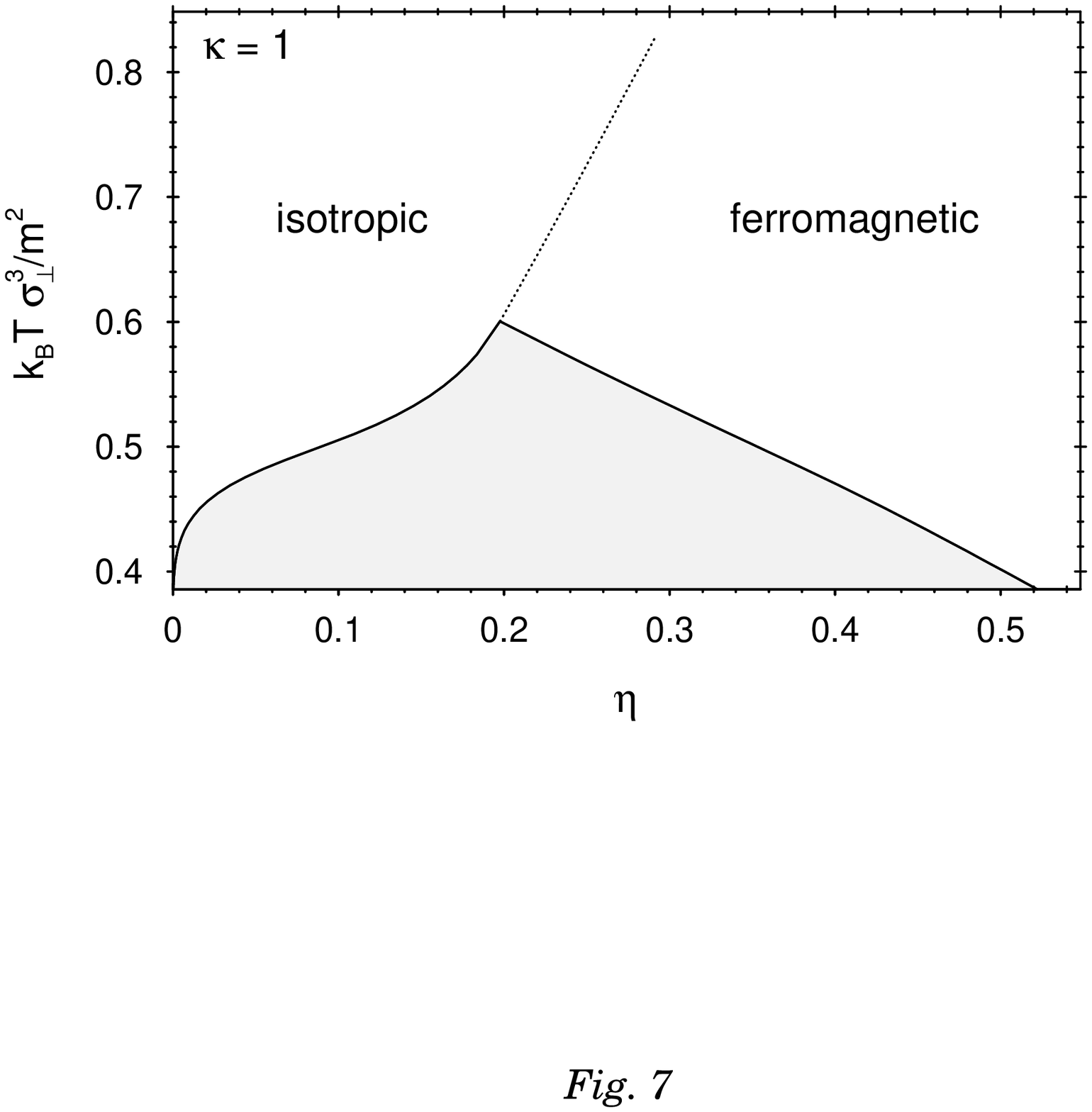}
\end{minipage}

\newpage
\begin{minipage}[h]{16cm}
\pssize{16cm}
\epsfbox{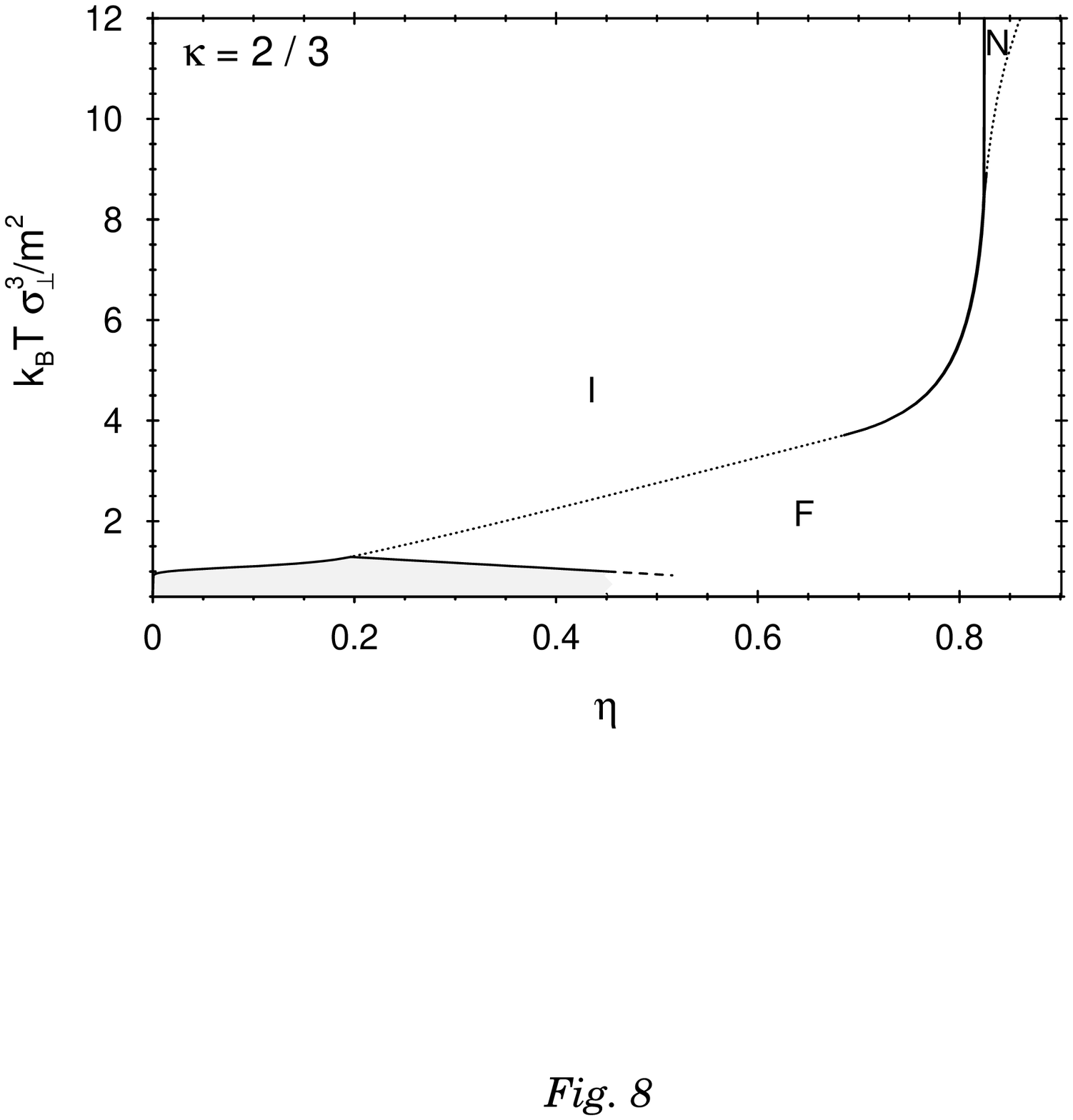}
\end{minipage}

\newpage
\begin{minipage}[h]{16cm}
\pssize{16cm}
\epsfbox{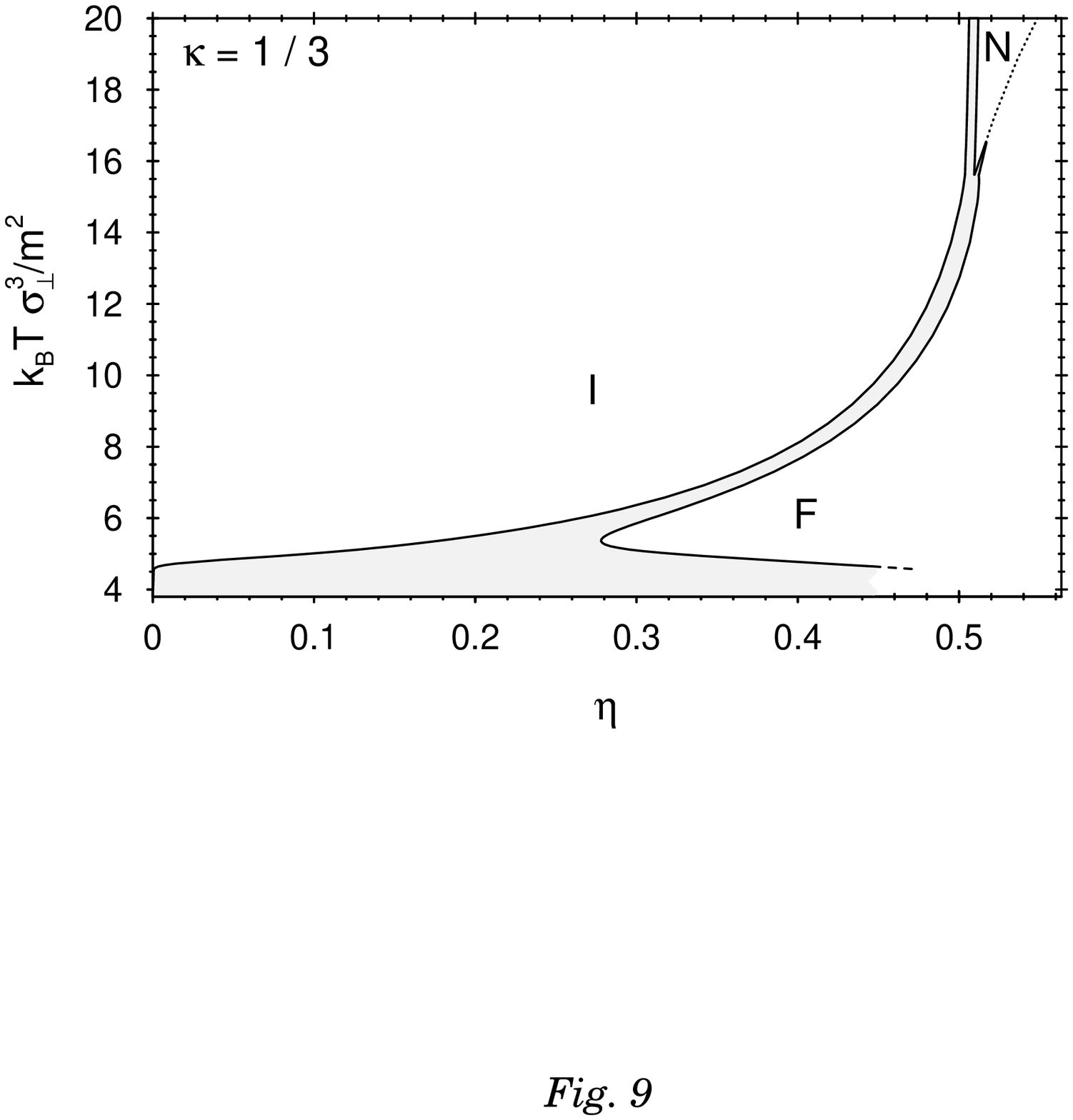}
\end{minipage}

\newpage
\begin{minipage}[h]{16cm}
\pssize{16cm}
\epsfbox{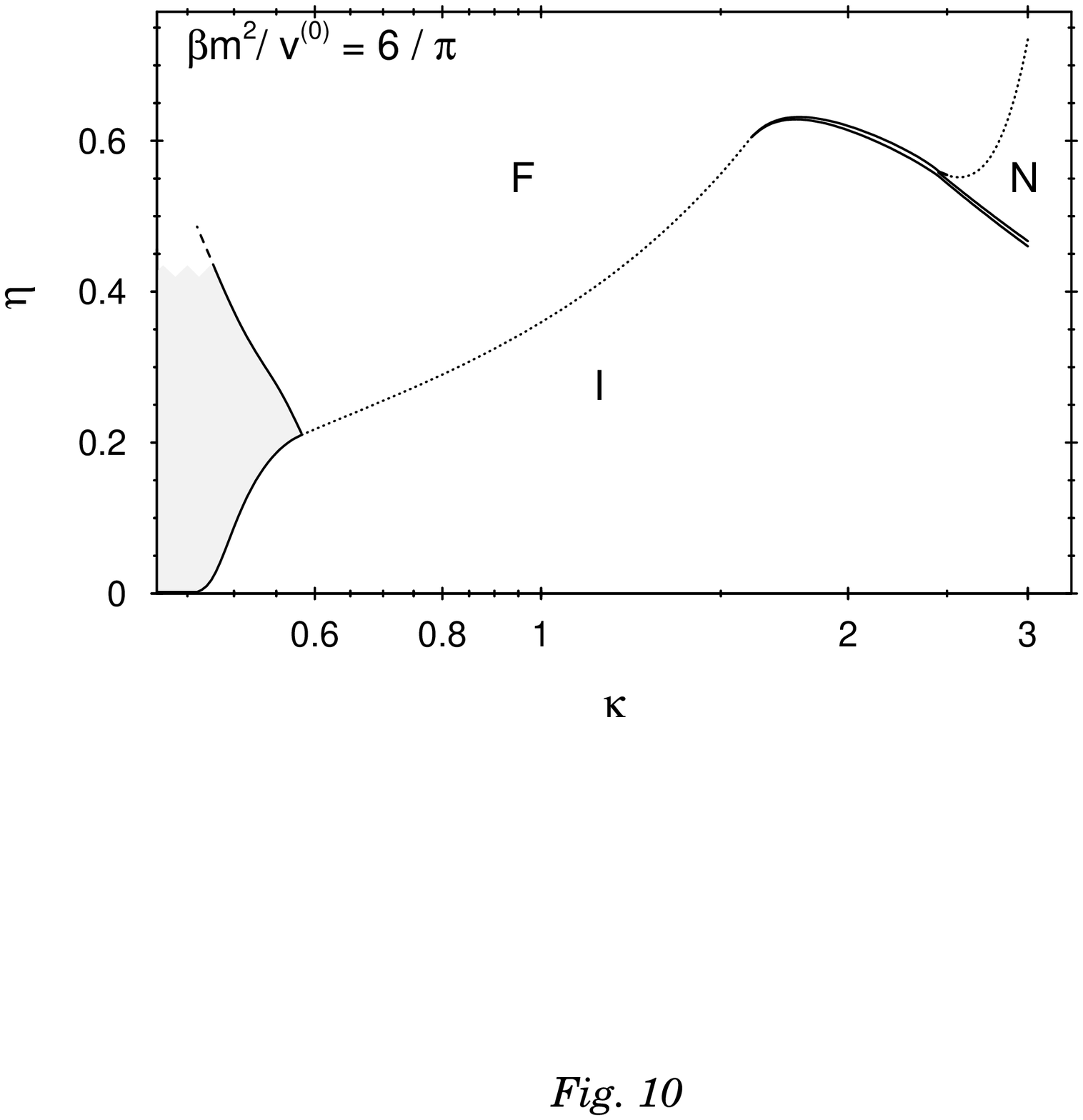}
\end{minipage}

\newpage
\begin{minipage}[h]{16cm}
\pssize{16cm}
\epsfbox{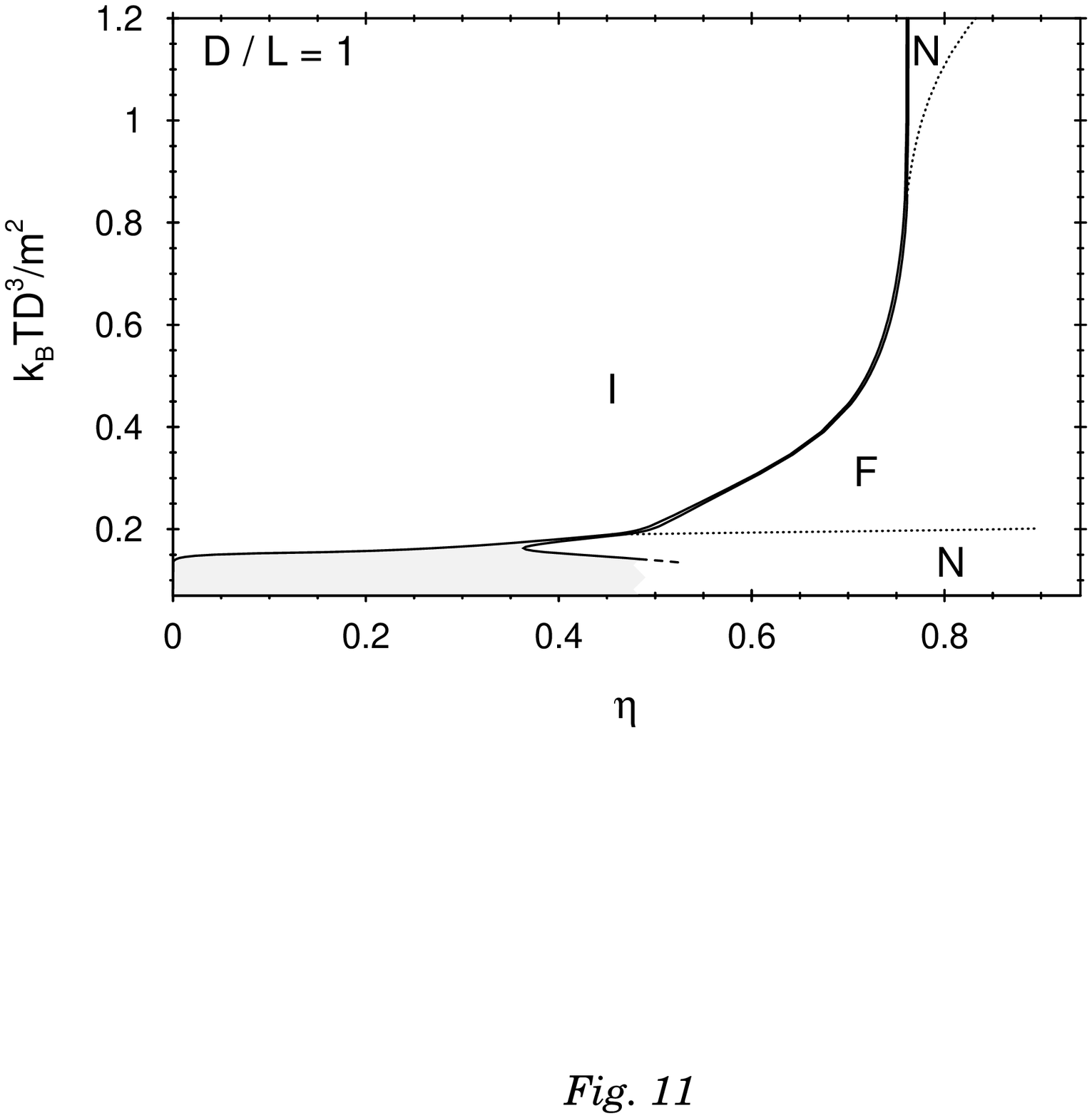}
\end{minipage}


\references

\bibitem{Onsager} L.~Onsager, Proc. NY. Acad. Sci. {\bf 51}, 627
(1949).

\bibitem{LesHouches} D.~Frenkel, in {\it Les Houches Summer School
Lectures, Session LI}, edited by J.P.~Hansen, D.~Levesque, and
J.~Zinn-Justin  (Elsevier, Amsterdam, 1991), p. 689.

\bibitem{TarazonaRev} P.~Tarazona, Phil. Trans. R. Soc. Lond. A {\bf
344}, 307 (1993).

\bibitem{AllenRev} M.P.~Allen, G.T.~Evans, D.~Frenkel, and
B.M.~Mulder, Adv. Chem. Phys. {\bf 86}, 1 (1993).

\bibitem{FrenkelEll} D.~Frenkel and B.M.~Mulder, Mol. Phys. {\bf 55},
1171 (1985).

\bibitem{FrenkelNat} D.~Frenkel, H.N.W.~Lekkerkerker, and
A.~Stroobants, Nature {\bf 332}, 822 (1988).

\bibitem{Frenkel88} D.~Frenkel, J. Phys. Chem. {\bf 92}, 3280 (1988).

\bibitem{Veerman} J.A.C.~Veerman and D.~Frenkel, Phys. Rev. A {\bf
41}, 3237 (1990).

\bibitem{McGrotherHSC} S.C.~McGrother, D.C.~Williamson, and
G.~Jackson, J. Chem. Phys. {\bf 104}, 6755 (1996).

\bibitem{FrenkelCS} D.~Frenkel, Liq. Cryst. {\bf 5}, 929 (1989).

\bibitem{VeermanCS} J.A.C.~Veerman and D.~Frenkel, Phys. Rev. A {\bf
45}, 5633 (1992).

\bibitem{Patey1} D.~Wei and G.N.~Patey, Phys. Rev. Lett. {\bf 68},
2043 (1992).

\bibitem{Patey2} D.~Wei and G.N.~Patey, Phys. Rev. A {\bf 46}, 7783
(1992).

\bibitem{Weis1} J.J.~Weis, D.~Levesque, and G.J. Zarragoicoechea,
Phys. Rev. Lett. {\bf 69}, 913 (1992).

\bibitem{Weis2} J.J.~Weis and D.~Levesque, Phys. Rev. E {\bf 48}, 3728
(1993).

\bibitem{Grest2} M.J.~Stevens and G.S.~Grest, Phys. Rev. E {\bf 51},
5962 (1995).

\bibitem{Grest} M.J.~Stevens and G.S.~Grest, Phys. Rev. E {\bf 51},
5976 (1995).

\bibitem{Patey3} D.~Wei, G.N.~Patey, and A.~Perera, Phys. Rev. E {\bf
47}, 506 (1993).

\bibitem{Letter} B.~Groh and S.~Dietrich, Phys. Rev. Lett. {\bf 72},
2422 (1994); ibid {\bf 74}, 2617 (1995).

\bibitem{Paper} B.~Groh and S.~Dietrich, Phys. Rev. E {\bf 50}, 3814
(1994).

\bibitem{Domains} B.~Groh and S.~Dietrich, Phys. Rev. E {\bf 53}, 2509 (1996).

\bibitem{Griechen} A.G.~Vanakaras and D.J.~Photinos, Mol. Phys. {\bf
85}, 1089 (1995).

\bibitem{PateyEll} A.~Perera and G.N.~Patey, J. Chem. Phys. {\bf 91},
3045 (1989).

\bibitem{BausColot} M.~Baus and J.-L.~Colot, Phys. Rev. A {\bf 40},
5444 (1989).

\bibitem{Vega} C.~Vega and S.~Lago, J. Chem. Phys. {\bf 100}, 6727
(1994).

\bibitem{ZLW92} G.J.~Zarragoicoechea, D.~Levesque, and J.J.~Weis,
Mol. Phys. {\bf 75}, 989 (1992).

\bibitem{WLZ93} J.J.~Weis, D.~Levesque, and G.J. Zarragoicoechea,
Mol. Phys. {\bf 80}, 1077 (1993).

\bibitem{LWZ93} D.~Levesque, J.J.~Weis and G.J. Zarragoicoechea,
Phys. Rev. E {\bf 47}, 496 (1993).

\bibitem{Satoh} K.~Satoh, S.~Mita, and S.~Kondo, Liq. Cryst. {\bf 20},
757 (1996).

\bibitem{Terentjev} E.M.~Terentjev, M.A.~Osipov, and T.J.~Sluckin,
J. Phys. A: Math. Gen. {\bf 27}, 7047 (1994).

\bibitem{Colot} J.-L.~Colot, X.-G. Wu, H.~Xu, and M.~Baus,
Phys. Rev. A {\bf 38}, 2022 (1988).

\bibitem{Pynn} R.~Pynn, Solid State Commun. {\bf 14}, 29 (1974);
J. Chem. Phys. {\bf 60}, 4579 (1974).

\bibitem{Hansen} J.P.~Hansen and I.R. MacDonald,
{\it Theory of Simple Liquids} (Academic, London, 1976).

\bibitem{Lee1} S.-D.~Lee, J. Chem. Phys. {\bf 87}, 4972 (1987).

\bibitem{Lee2} S.-D.~Lee, J. Chem. Phys. {\bf 89}, 7036 (1988).

\bibitem{CS} N.F.~Carnahan and K.E.~Starling, J. Chem. Phys. {\bf
51}, 635 (1969).

\bibitem{Telo} P.I.~Teixeira and M.M.~Telo da Gama,
J. Phys.: Condens. Matter {\bf 3}, 111 (1991).

\bibitem{Frodl1} P.~Frodl and S.~Dietrich, Phys. Rev. A {\bf 45}, 7330
(1992); Phys. Rev. E {\bf 48}, 3203 (1993).

\bibitem{TeloHeisenberg} J.M.~Tavares, M.M.~Telo da Gama,
P.I.C.~Teixeira, J.J.~Weis, and M.J.P.~Nijmeijer, Phys. Rev. E {\bf
52}, 1915 (1995).

\bibitem{Griffiths} R.B.~Griffiths, Phys. Rev. {\bf 176}, 655 (1968).

\bibitem{Chandrasekhar} S.~Chandrasekhar, {\it Liquid Crystals}
(Cambridge University Press, Cambridge, 1992).

\bibitem{Berne} B.J.~Berne and P.~Pechukas, J. Chem. Phys. {\bf 56},
4213 (1972).

\bibitem{Solid} B.~Groh and S.~Dietrich, Phys. Rev. E {\bf 54}, 1687
(1996).

\bibitem{McGrother} S.C.~McGrother and G.~Jackson,
Phys. Rev. Lett. {\bf 76}, 4183 (1996).

\bibitem{Smit} M.E. van Leeuwen and B.~Smit, Phys. Rev. Lett. {\bf
71}, 3991 (1993).

\bibitem{RoijLett} R. van Roij, Phys. Rev. Lett. {\bf 76}, 3348
(1996).

\bibitem{Roij} R. van Roij, P.~Bolhuis, B.~Mulder, and D.~Frenkel,
Phys. Rev. E {\bf 52}, R1277 (1995).
\end{document}